\documentclass[preprint, 3p, sort&compress]{elsarticle}

\usepackage{amssymb}
\usepackage{amsthm}
\usepackage[mathlines]{lineno}
\usepackage{graphicx}
\usepackage{units}
\usepackage{url}
\usepackage{amsmath}
\usepackage{amsfonts}
\usepackage{bm}
\usepackage{textcomp}
\usepackage{subfigure}
\usepackage{multicol}
\usepackage{multirow}
\usepackage{verbatim}
\usepackage{rotating}
\usepackage[colorlinks,linkcolor=black,citecolor=red]{hyperref}
\usepackage{float}
\usepackage{epsfig}
\usepackage{dcolumn}
\usepackage{bm}
\usepackage{color}
\usepackage{pstricks}
\usepackage{pst-node}
\usepackage{times}
\usepackage{indentfirst}
\usepackage[english]{babel}
\usepackage{epstopdf}
\addto{\captionsenglish}{%

}
\journal{Physics Letters B}


\usepackage[font=footnotesize]{caption}

\usepackage{overpic}

\newcommand{\ee}{e^+e^-}
\newcommand{\ra}{\rightarrow}

\biboptions{numbers,sort&compress}

\begin{document}

\begin{frontmatter}

\title{\boldmath Measurements of $\Sigma^{+}$ and $\Sigma^{-}$ Time-like Electromagnetic Form Factors for center-of-mass energies from $2.3864$ to $3.0200$~GeV}

\author{
M.~Ablikim$^{1}$, M.~N.~Achasov$^{10,d}$, P.~Adlarson$^{64}$, S.~Ahmed$^{15}$, M.~Albrecht$^{4}$, A.~Amoroso$^{63A,63C}$, Q.~An$^{60,48}$, ~Anita$^{21}$, Y.~Bai$^{47}$, O.~Bakina$^{29}$
% ...
\\
\vspace{0.2cm}
(BESIII Collaboration)\\
\vspace{0.2cm} 
$^{1}$ Institute of High Energy Physics, Beijing 100049, People's Republic of China\\
% ...
$^{4}$ Bochum Ruhr-University, D-44780 Bochum, Germany\\
% ...
$^{10}$ G.I. Budker Institute of Nuclear Physics SB RAS (BINP), Novosibirsk 630090, Russia\\
% ...
$^{15}$ Helmholtz Institute Mainz, Johann-Joachim-Becher-Weg 45, D-55099 Mainz, Germany\\
% ...
$^{21}$ Indian Institute of Technology Madras, Chennai 600036, India\\
% ...
$^{29}$ Joint Institute for Nuclear Research, 141980 Dubna, Moscow region, Russia\\
% ...
$^{47}$ Southeast University, Nanjing 211100, People's Republic of China\\
$^{48}$ State Key Laboratory of Particle Detection and Electronics, Beijing 100049, Hefei 230026, People's Republic of China\\
% ...
$^{60}$ University of Science and Technology of China, Hefei 230026, People's Republic of China\\
% ...
$^{63}$ (A)University of Turin, I-10125, Turin, Italy; (B)University of Eastern Piedmont, I-15121, Alessandria, Italy; (C)INFN, I-10125, Turin, Italy\\
$^{64}$ Uppsala University, Box 516, SE-75120 Uppsala, Sweden\\
% ...
$^{d}$ Also at the Novosibirsk State University, Novosibirsk, 630090, Russia\\
% ...
\vspace{0.4cm}
}

\date{\today}

\begin{abstract}
The Born cross sections of the $e^{+}e^{-}\to\Sigma^{+}\bar{\Sigma}^{-}$ and $e^{+}e^{-}\to\Sigma^{-}\bar{\Sigma}^{+}$ processes 
are determined for center-of-mass energy from 2.3864 to 3.0200~GeV with the BESIII detector.
The cross section lineshapes can be described properly by a pQCD function and  
the resulting ratio of effective form factors for the $\Sigma^{+}$ and $\Sigma^{-}$ is consistent with 3. 
In addition, ratios of the $\Sigma^{+}$ electric and magnetic form factors, $|G_{E}/G_{M}|$, 
are obtained at three center-of-mass energies through an analysis of the angular distributions.
These measurements, which are studied for the first time in the off-resonance region, provide precision experimental input for understanding baryonic structure.
The observed new features of the $\Sigma^{\pm}$ form factors require more theoretical discussions for the hyperons. 
\end{abstract}

\end{frontmatter}

\begin{multicols}{2}
\section{Introduction}
Nucleons, as the lightest baryons, are the largest component of the observable matter in the universe, and
were shown to be non-pointlike particles in the middle of last century~\cite{stern,hofstadter}. 
However, nucleon properties, such as their radii and the sources of their spin, are still not well
understood~\cite{protonradius}. 
The hyperons are the $SU(3)$-flavour-octet partners of the nucleons that contain one or more strange quarks,
and offer crucial additional dimensions to the study of nucleon structures~\cite{hyperon1,hyperon2}.
Treating the heavier strange quarks as spectators, hyperons can provide valuable insight into the behaviour of the lighter up and down quarks in different environments.
Electromagnetic form factors~(EMFFs) are fundamental observables of baryons
that are intimately related to their internal structure and dynamics~\cite{chiral, pqcd, lattice1}.
Despite the fact that much work has been done on the EM structures of protons in both the space-like and 
time-like regions~\cite{space_ros, space_jlab, time_cleo, time_babar, time_cmd, time_bes3},
experimental information regarding the EMFFs of hyperons remains limited~\cite{hyperon_babar, hyperon_cleo, hyperon_bes3, hyperon_bes32}.
Moreover, the few existing measurements of time-like neutron FFs~\cite{nnbar_fenice, nnbar_bes3} differ from
each other and lead to conflicting conclusions when compared to those for the proton~\cite{ratio_theory1, ratio_theory2}. 
A $\Sigma^+$ hyperon is formed by replacing the proton's down quark with a strange quark; likewise a
$\Sigma^-$ is formed by replacing  the neutron's up quark with a strange quark.
The corresponding ratio of FFs between  the $\Sigma^{+}$ and $\Sigma^{-}$ hyperons could provide 
guidance for the nucleons.
Therefore, experimental measurements for $\Sigma$ hyperons,
especially the $\Sigma^{-}$, which has never been measured in the time-like region, provide essential  
tests of various theoretical models~\cite{ratio_theory2,diquark, UandA} and produce important input for the understanding of baryonic structures.

The differential, one-photon exchange cross section for the $\ee\ra B\bar{B}$ process,
where $B$ is a spin-1/2 baryon, can be expressed in terms of the electric and magnetic FFs $G_{E}$ and $G_{M}$ 
as~\cite{ee}:
\begin{equation}
\label{equ:eq1}
\frac{d\sigma^{B}(s)}{d\Omega} = \frac{\alpha^{2}\beta C|G_{M}|^{2}}{4s}\left[(1+\cos^{2}\theta) +
  \frac{1}{\tau}\left|\frac{G_{E}}{G_{M}}\right|^{2}\sin^{2}\theta\right],
\end{equation}
where $\alpha$ is the fine-structure constant, 
$s$ is the square of center-of-mass~(c.m.)~energy,
$\beta=\sqrt{1-4m_{B}^{2}/s}$ is a phase-space factor, 
$\tau=\frac{s}{4m_{B}^{2}}$,
$m_{B}$ is the baryon mass, and $\theta$ is its c.m.~production angle.
The Coulomb correction factor $C$~\cite{schwinger,coulomb}
accounts for the electromagnetic interaction of charged point-like fermion pairs in the final state.
It reads  $C=y/(1-e^{-y})$ with $y=\pi\alpha(1+\beta^{2})/\beta$
for a charged point-like fermion pair and $C=1$ for a neutral point-like fermion pair.
For charged point-like fermion pairs, the cross section at threshold is non-zero,
$\sigma(4m^{2}_B)=\pi^2\alpha^3/2m^2_B = 848(m_p/m_B)^2$~pb, where $m_{p}$ is the proton mass~\cite{pdg},
and then grows with increasing $\beta$.
Experimentally, a rapid rise of the $e^{+}e^{-}\rightarrow p\bar{p}$ cross section near threshold
followed by a plateau is observed~\cite{time_babar, time_cmd}. 
The crosss section of plateau near threshold is consistent with the 848~pb expectation for a point-like charged particle.
However, in this case, the $p\bar{p}$ is produced by a virtual photon with $Q^2=4m_p^2=3.53$~GeV$^2$,
which corresponds to a Compton wavelength of $\sim$0.1~fm,
a scale at which the proton is definitely not point-like.
A similar feature of the cross section for $e^{+}e^{-}\to \Lambda_{c}^{+}\bar{\Lambda}_{c}^{-}$ is observed by the BESIII experiment~\cite{lambdac_bes3},
where the cross section of plateau near threshold is around 240~pb. This is 
1.6 times the predicted value for point-like charged particles. 
These unexpected threshold effects have been widely discussed in the literature 
where they are interpreted as
final state interactions~\cite{int_fsi}, bound states or near-threshold meson resonances~\cite{int_res}, or an attractive Coulomb interaction~\cite{int_coulomb}.
To understand the nature of these threshold effects, experimental measurements of the near threshold charged pair production of other hyperons will be of critical importance.

\section{Detector and data sample}
In this Letter, we present precision measurements of $e^{+}e^{-}\to\Sigma^{+}\bar{\Sigma}^{-}$
and $e^{+}e^{-}\to\Sigma^{-}\bar{\Sigma}^{+}$ with a data sample of 329.7~pb$^{-1}$
collected at BESIII with c.m.~energies between $2.3864$ and 3.0200~GeV~\cite{dataset}.
The threshold energies for $\Sigma^{+}\bar{\Sigma}^{-}$ and $\Sigma^{-}\bar{\Sigma}^{+}$ pair production are 
$2.3787$~GeV and 2.3949~GeV, respectively.
The BESIII detector is described in detail in Ref.~\cite{Ablikim:2009aa}. 
The critical
elements for the measurements reported here are: the main drift chamber (MDC), which
measures the momenta of charged particles with 0.5\% resolution for 1~GeV/$c$ tracks and
the $dE/dx$ for charged-particle identification (PID); a barrel array of scintillation counters that
measures charged particles' time of flight for additional PID information; and an electromagnetic
calorimeter (EMC) comprising an array of CsI(Tl) crystals that measures photon energies with a
resolution of 2.5\% at 1 GeV. 

Simulated event samples produced with a {\sc geant4}-based~\cite{geant4} Monte Carlo~(MC)
package that includes the geometric description of the BESIII detector and its response,
are used to determine the detection efficiency and to estimate the backgrounds.
The signal processes $e^{+}e^{-}\to\Sigma^{\pm}\bar{\Sigma}^{\mp}$ are generated according to 
the differential amplitude presented in Ref.~\cite{Anjrej}.
Initial state radiation~(ISR) is simulated with {\sc conexc}~\cite{conexc} and the corresponding correction factors
are calculated for higher order processes.
 Background from the QED processes $e^{+}e^{-}\to l^{+}l^{-}~(l=e,\mu)$ and $e^{+}e^{-}\to\gamma\gamma$  are investigated  with {\sc babayaga}~\cite{babayaga},
while for $e^{+}e^{-}\to $hadrons and two-photon processes we use
{\sc lundarlw}~\cite{lundarlw} and {\sc bestwogam}~\cite{bestwogam}, respectively.

\section{Data Analysis}
In the process $e^{+}e^{-}\to\Sigma^{+}\bar{\Sigma}^{-}$,  there are four dominant final state topologies
which account for more than 99\% of its total decay width:
\noindent 
$p\pi^{0}\bar{p}\pi^{0}$, $n\pi^{+}\bar{p}\pi^{0}$, $p\pi^{0}\bar{n}\pi^{-}$ and $n\pi^{+}\bar{n}\pi^{-}$,
All four configurations are selected in this analysis, significantly improving the statistics.
At BESIII, charged particles are efficiently detected and identified by the MDC and PID systems and
$\pi^{0}$ mesons are reconstructed in the EMC via their $\pi^0\to\gamma\gamma$ decay mode. 
The selection criteria for charged tracks, PID, and photon candidates are the same as those used
in Ref.~\cite{pinglamlam}.  
Most of the anti-neutrons ($\bar{n}$) annihilate in the EMC and produce several
secondary particles with a total energy deposition that can be as high as 2~GeV; 
the position of the $\bar{n}$ interaction and, from this,
the $\bar{n}$ direction can be inferred from the weighted center-of-energy of the shower~\cite{hyperon_bes3}. 
Neutron~($n$)  detection is not done because of its low interaction efficiency and small energy deposition.

The $p\pi^{0}\bar{p}\pi^{0}$ and $n\pi^{+}\bar{p}\pi^{0}$ final-state configurations, classified
as category A, can be analyzed by a partial reconstruction technique in which only the detection of
$\bar{\Sigma}^{-}\to\bar{p}\pi^{0}$ is required.
Candidate events are required to have at least one charged track that is identified as a $\bar{p}$ by the PID
system  and at least two good photons that are consistent with originating from $\pi^0\to\gamma\gamma$.
The mass spectrum of $\gamma\gamma$ is required to be from $0.127<M_{\gamma\gamma}<0.139$~GeV/c$^{2}$
to $0.123<M_{\gamma\gamma}<0.14$~GeV/c$^{2}$, depending on c.m.~energies. The $\bar{\Sigma}^{-}$ is reconstructed using all combinations of the 
selected $\bar{p}\gamma\gamma$.
The two-body process exploits two variables that are based on energy and momentum conservation: the energy difference 
$\Delta E\equiv E-E_{\rm beam}$ and the beam-constrained mass $M_{\text{bc}}\equiv\sqrt{E^{2}_{\text{beam}}-p^{2}}$.
Here, $E(p)$ is the total measurement energy (momentum) of the $\bar{p}\gamma\gamma$ combinations in the c.m.~system,
and $E_{\rm beam}$ is the beam energy.
Candidates are accepted with optimized $\Delta E$ requirements of $-16<\Delta E<7$~MeV to $-24<\Delta E<13$~MeV, depending on c.m.~energies, and with $M_{\text{bc}}>1.15$~GeV/c$^{2}$.

The $p\pi^{0}\bar{n}\pi^{-}$ and $n\pi^{+}\bar{n}\pi^{-}$ final states, classified as category B,
are reconstructed by requiring two good charged tracks with one identified as a $\pi^{-}$ and the other
identified as either a $\pi^{+}$ or $p$,
and the most energetic shower in these events is assigned as the
$\bar{n}$ candidate. To discriminate $\bar{n}$-initiated showers from those produced by photons, three variables
are retained for further selection based on c.m.~energy-dependent requirements: the total energy in the $\bar{n}$-assigned  EMC shower,
the second moment of the shower~\cite{hyperon_bes3}, and the 
number of crystals with above-threshold signals within a 40$^{\circ}$ cone around the shower.
After that, kinematic fits that include the $\bar{n}$ direction are performed to identify signal events.
Since the $\bar{n}$ shower does not provide a good measure of its total energy, $E_{\bar{n}}$, this is left
as a free parameter in the kinematic fits. 
If a $\pi^{+}$ is identified, the fit imposes the $n\bar{n}\pi^{+}\pi^{-}$ hypothesis with a missing $n$.
If a $p$ is identified, the fit imposes the $p\bar{n}\pi^{-}\pi^{0}$ hypothsis with a missing $\pi^{0}$.
In both fits, total energy-momentum conservation is constrained and $M_{\bar{n}\pi^{-}}$ is also constrained to the mass of the $\bar{\Sigma}^{-}$.
The $p\pi^{-}$ invariant mass is required
to be  $|M(p\pi^{-})-m(\Lambda)|>0.005$~GeV/$c^{2}$ to eliminate background from
$e^{+}e^{-}\to\Lambda\bar{\Lambda}\to p\pi^{-}\bar{n}\pi^{0}$.
Furthermore, the $\chi^{2}$ value from the kinematic fit is required to be less than 20.

The reconstruction of $e^{+}e^{-}\to\Sigma^{-}\bar{\Sigma}^{+}$ is similar to that for $n\pi^{+}\bar{n}\pi^{-}$ 
in the $e^{+}e^{-}\to\Sigma^{+}\bar{\Sigma}^{-}$ analysis since they have the same final states. 
The only difference is that
$M_{\bar{n}\pi^{+}}$ is constrained to the mass of the $\bar{\Sigma}^{+}$ in the kinematic fit.

\begin{figure*}[htbp]
\begin{center}
\begin{overpic}[width=8.cm,height=3.5cm,angle=0]{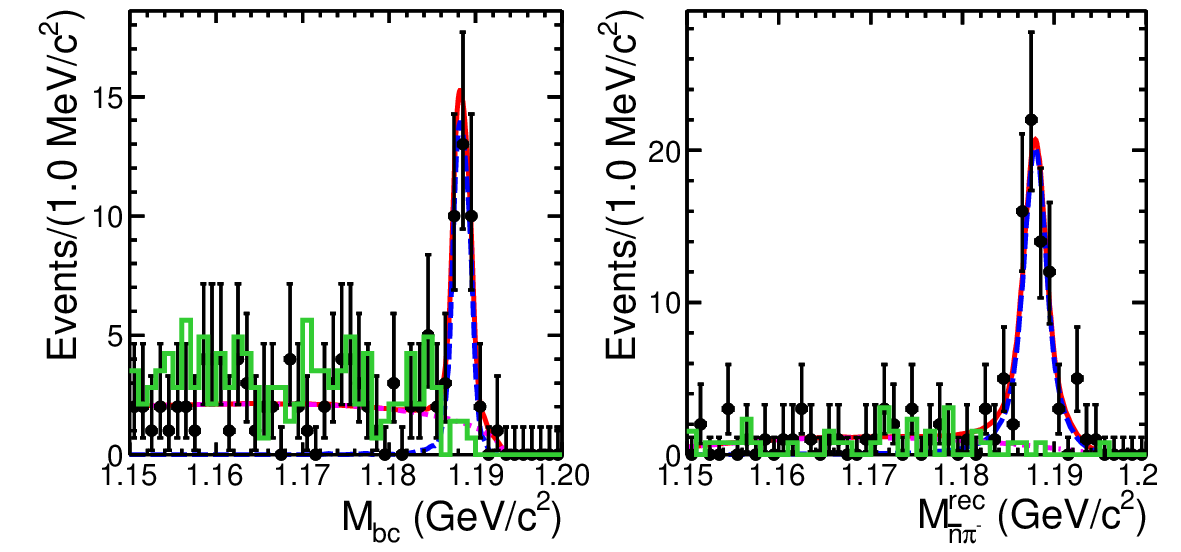}
\put(14,33){\small{a)}}
\put(61,33){\small{b)}}
\put(14,24){\tiny{$\sqrt{s}=2.3864$~GeV}}
\put(63,24){\tiny{$\sqrt{s}=2.3864$~GeV}}
\end{overpic}
\begin{overpic}[width=8.cm,height=3.5cm,angle=0]{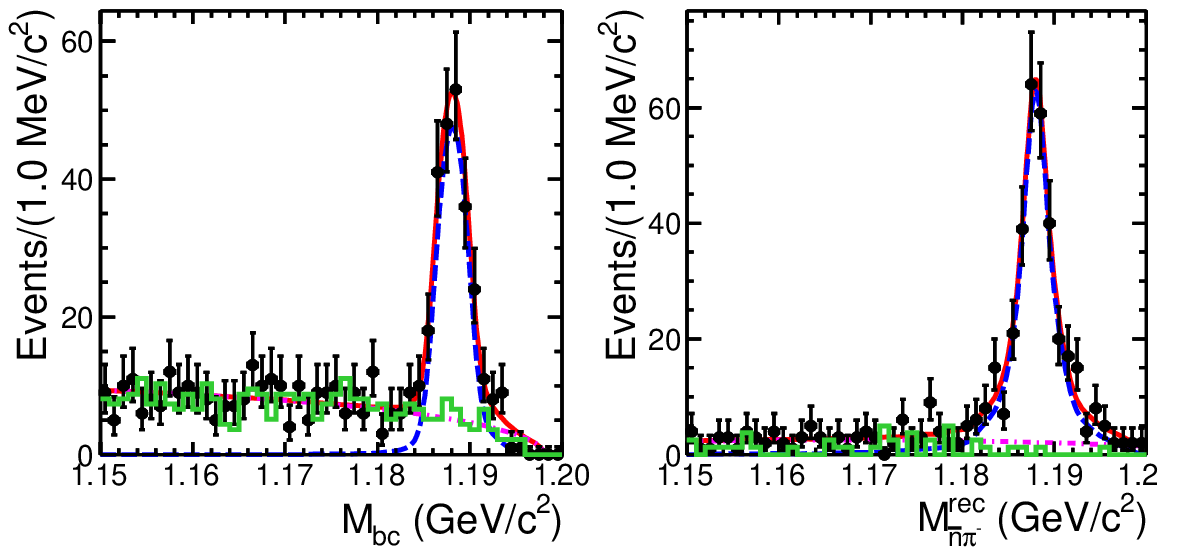}
\put(14,34){\small{c)}}
\put(61,34){\small{d)}}
\put(14,24){\tiny{$\sqrt{s}=2.3960$~GeV}}
\put(63,24){\tiny{$\sqrt{s}=2.3960$~GeV}}
\end{overpic}
\caption{ (color online) 
The mass spectra of $M_{\text{bc}}$ (category A) and $M^{\rm rec}_{\bar{n}\pi^{-}}$ (category B) 
for $e^{+}e^{-}\to\Sigma^{+}\bar{\Sigma}^{-}$ candidate events 
at a,b) $\sqrt{s}=2.3864$~GeV. 
Dots with error bars are the data; histograms are the background events in
MC samples after normalization. Solid curves are the fit results,
dashed curves are the signal, and  dot-dashed curves are the background.}
\vspace{-0.5cm}
\label{stacksigmap}
\end{center}
\end{figure*}

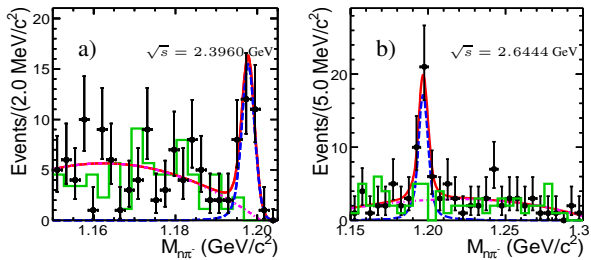
\begin{figure}[H] 
\begin{center}
\begin{overpic}[width=8.cm,height=3.5cm,angle=0]{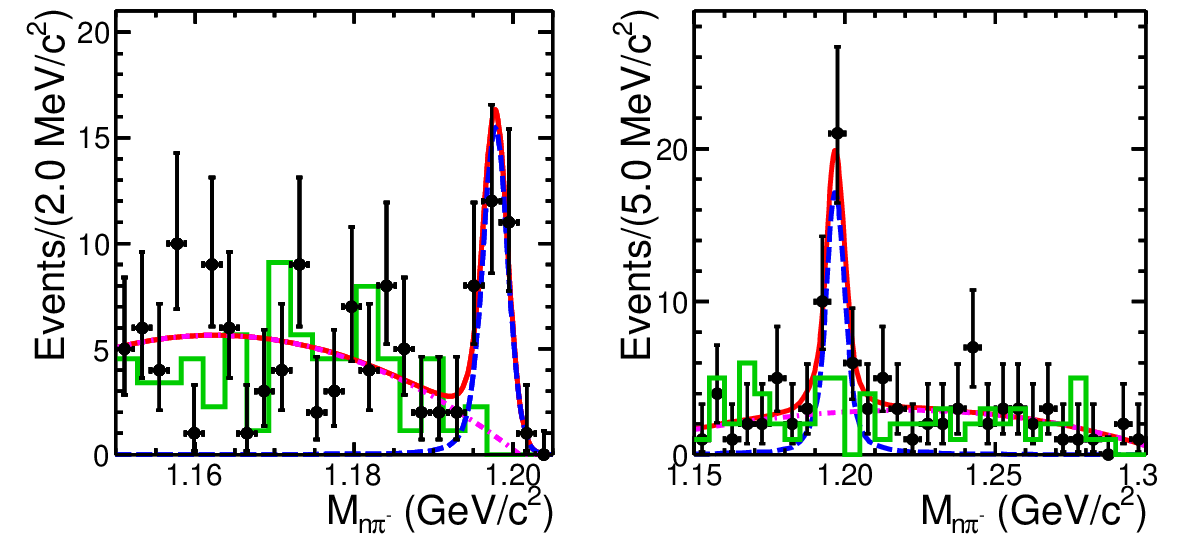}
\put(14,35){\small{a)}}
\put(63,35){\small{b)}}
\put(25,35){\tiny{$\sqrt{s}=2.3960$~GeV}}
\put(75,35){\tiny{$\sqrt{s}=2.6444$~GeV}}
\end{overpic}
\caption{ (color online) The $M_{n\pi^{-}}$ distributions for selected $e^{+}e^{-}\to\Sigma^{-}\bar{\Sigma}^{+}$
events at a) $\sqrt{s}=2.3960$~GeV and b) $\sqrt{s}=2.6444$~GeV.
Dots with error bars are the data; histograms are the background events in
MC after normalization. Solid curves are the fit results,
 dashed curves are the signal, and dot-dashed curves are the background.}
\vspace{-0.5cm}
\label{stacksigmam}
\end{center}
\end{figure}

Figure~\ref{stacksigmap} 
shows the distributions of $M_{bc}$ for category A and the recoil mass of $\bar{n}\pi^{-}$, $M^{\rm rec}_{\bar{n}\pi^{-}}$, 
for category B using selected $e^{+}e^{-}\to\Sigma^{+}\bar{\Sigma}^{-}$ candidates, 
where significant signals in both categories are observed in data at $\sqrt{s}=2.3864$ and
2.3960~GeV. 
Backgrounds are studied with MC
samples and only hadronic final states survive the selection criteria. In category~A,
the backgrounds are from $e^+e^-$ annihilation events with the same final states as the signal
process, with one or more additional $\pi^{0}$, and with an additional $\gamma$-ray.
In category~B,  the backgrounds are from annihilation events with the same final states as the
signal process, multi-$\pi$ processes such as $\pi^{+}\pi^{-}\pi^{0}\pi^{0}$ and processes with one more
$\pi^{0}$ in the final states.  
These background processes are mainly from contributions including intermediate states such as $\Delta$,
$\Lambda$ and $\Sigma$ baryons, but none of them produce peaks in the signal regions as shown by the histograms of  Fig.~\ref{stacksigmap}.
Figure~\ref{stacksigmam} shows distributions of $M_{n\pi^{-}}$ for
$e^{+}e^{-}\to\Sigma^{-}\bar{\Sigma}^{+}$ candidate events at $\sqrt{s}=2.3960$ and
2.6444~GeV, respectively, where significant signals in data are observed. 
In the background study, no peaking background is observed in the $n\pi^{-}$ mass spectrum.

\end{multicols}

\begin{multicols}{2}

The Born cross section for $e^{+}e^{-}\to\Sigma^{+}\bar{\Sigma}^{-}$ is determined from the relation:
\begin{equation}
\label{calsigma}
\sigma^{\text{B}} = \frac{N_{i}}{\mathcal{L}(1+\delta^{r})\frac{1}{|1-\Pi|^{2}}\delta^{\text{data/MC}}_{i}\mathcal{B}_{i}\varepsilon_{i}},~(i=A,B),
\end{equation} 
where $N$ is the signal yield extracted from the fits; 
$\mathcal{L}$ is the integrated luminosity; 
$1+\delta^{r}$ is the ISR correction factor incorporating the input cross section from 
this analysis iteratively;
$\frac{1}{|1-\Pi|^{2}}$ is the vacuum polarization factor~\cite{VP};
$\varepsilon$ is the detection efficiency determined from signal MC events.
The factor $\delta^{\text{data/MC}}$ is a correction factor for efficiency differences between data and MC simulation,
determined from studies of high statistics, low-background control samples of
$J/\psi\to\Sigma^{+}\bar{\Sigma}^{-}$ and $J/\psi\to\Lambda\bar{\Sigma}^{-}\pi^{+}$, respectively.
The decay branching fraction $\mathcal{B}$ accounts for the intermediate states in the $\bar{\Sigma}^{-}$ decay
(51.57\% for $\bar{\Sigma}^{-}\to \bar{p}\pi^{0}$ and  48.31\% for $\bar{\Sigma}^{-}\to \bar{n}\pi^{-}$).

To determine the signal yields, un-binned maximum likelihood fits are performed to the $M_{\text{bc}}$ and $M_{n\pi^{+}}$
distributions for categories A and B, respectively.
The probability density function (PDF) for the signal is described with a MC-simulated shape
convolved with a Gaussian function to account for mass resolution differences between
data and MC simulation.
The background PDF for category A is described by an Argus function~\cite{argus};
for category B by a second order polynomial. 
In the fit, the two categories are constrained by the same Born cross section $\sigma^{\text{Born}}$,
and the expected signal yields are calculated from
$N_{i}=\sigma^{\text{Born}}\cdot \mathcal{L}\cdot \varepsilon_{i}\cdot (1+\delta)\cdot \delta^{\text{data/MC}}_{i}\cdot \mathcal{B}_{i}$.
The fit results at $\sqrt{s}=2.3864$ and $\sqrt{s}=2.3960$~GeV are shown in Fig.~\ref{stacksigmap}.
Similarly, the signal yield of $e^{+}e^{-}\to\Sigma^{-}\bar{\Sigma}^{+}$ is determined by fitting 
the $n\pi^{-}$ mass spectrum, where the signal is described with the MC simulated shape convolved with a Gaussian
 function and the background is described with a 2nd-order polynomial.
Fit results at $\sqrt{s}=2.3960$ and $\sqrt{s}=2.6444$~GeV are shown in Fig.~\ref{stacksigmam}.

The quantities used in the cross section calculations for $e^{+}e^{-}\to\Sigma^{+}\bar{\Sigma}^{-}$ and $e^{+}e^{-}\to\Sigma^{-}\bar{\Sigma}^{+}$ are summarized in Table~\ref{sumsigmap} and Table~\ref{sumsigmam}, respectively.
It should be noted that, due to limited statistics, 
data at c.m~energies 2.7000 and 2.8000~GeV are combined;
data at   2.9500, 2.9810, 3.0000
and 3.0200 GeV are combined.
Currently, individual measurements on $|G_{E}|$ and $|G_{M}|$ at each energy point
are not possible due to statistics. Therefore, the effective FFs of $\Sigma^{\pm}$, defined as
$|G_{\text{eff}}|^2 \equiv (|G_E|^2+2\tau|G_M|^2)/(2\tau +1)$~\cite{effectiveff}, are reported here and shown in Table~\ref{sumsigmap},~\ref{sumsigmam}.
\end{multicols}

\begin{table*}[htbp]
\caption{Summary of the calculated cross section for $e^{+}e^{-}\to\Sigma^{+}\bar{\Sigma}^{-}$ and effective FFs of $\Sigma^{+}$ at each c.m.~energy and the quantities used in the
calculation, $\epsilon=\varepsilon(1+\delta^{r})\frac{1}{|1-\Pi|^{2}}\delta^{\text{data/MC}}$, defined in the text.
The energy points with asterisks are combined data samples with c.m~energies weighted by
the luminosities of the subsamples. The 2.7500~GeV is a combined data set of 2.7000 and 2.8000~GeV, and 2.9884~GeV is a combined data set of
 2.9500, 2.9810, 3.0000 and 3.0200~GeV.  The last column shows the results of $|G_{E}/G_{M}|$ ratio of $\Sigma^{+}$.
}
\vspace{-0.5cm}
\label{sumsigmap}
  \footnotesize
  \begin{center}
  \begin{tabular}{ c|c|ccccc}
  \hline
  $\sqrt{s}$~(GeV) & $\mathcal{L}$~(pb$^{-1}$) &  $\epsilon_{A}$(\%)  &  $\epsilon_{B}$(\%) &  $\sigma^{\text{Born}}$~(pb) & $|G_{\text{eff}}|$($\times10^{-2})$ &  $|G_{E}/G_{M}|$   \\
\hline
  2.3864	   & 22.6    	& 5.8  & 12.6  &  $58.2\pm5.9^{+2.8}_{-2.6}$& $16.5\pm0.9\pm0.9$   &-       \\
  2.3960		   & 66.9    	& 9.5 & 14.1 &  $68.6\pm3.4\pm2.3$     & $15.0\pm0.4\pm0.5$   &  $1.83\pm0.26\pm0.24$    \\
  2.5000		  & 1.10		    & 18.4 & 21.6 &  $130\pm29\pm11$	& $14.0\pm1.6\pm0.6$       &-        \\	
  2.6444	  & 33.7 	    & 24.4 & 20.5  &  $59.9\pm3.6\pm3.2$    &$8.6\pm0.3\pm0.2$      	& \multirow{2}{*}{$0.66\pm0.15\pm0.11$}  \\
  2.6464	& 34.0			& 24.2  & 20.7 &  $58.9\pm3.5\pm2.4$   &$8.5\pm0.3\pm0.2$       &          \\
  *2.7500	 & 2.04			& 25.0  & 19.7 &  $36.9\pm12.8\pm3.2$  & $6.7\pm1.2\pm0.3$       & -         \\
  2.9000   	& 105.	       & 26.5  & 20.6 &  $16.7\pm1.2\pm1.1$   & $4.5\pm0.2\pm0.2$       	&  $1.06\pm0.36\pm0.09$\\
  *2.9884	& 65.2			& 25.5  & 21.4 &  $12.4\pm1.3\pm1.3$    & $3.9\pm0.2\pm0.2$      & -       \\
  \hline
  \end{tabular}
  \end{center}
  \end{table*}

\begin{table*}[htbp]
\caption{Summary of the calculated cross section for $e^{+}e^{-}\to\Sigma^{-}\bar{\Sigma}^{+}$ and effective FFs of $\Sigma^{-}$ at each c.m.~energy and the quantities used in the
calculation.
}
\vspace{-0.5cm}
\label{sumsigmam}
  \footnotesize
  \begin{center}
  \begin{tabular}{ c|c|ccccc}
  \hline
  $\sqrt{s}$~(GeV) & $\mathcal{L}$~(pb$^{-1}$)  & $\epsilon$(\%)  &  $N$ & $\sigma^{\text{Born}}$~(pb) & $|G_{\text{eff}}|$($\times10^{-2})$  \\
\hline
  2.3864	   & 22.6       & \multicolumn{4}{c}{(below threshold)}\\
  2.3960		   & 66.9        & 18.8 & $29.6\pm6.7$ & $2.3\pm0.5\pm0.3$ & $3.9\pm0.5\pm0.6$\\
  2.5000		  & 1.10		        & 20.2  &$4.8^{+2.9}_{-2.2}$ & $21.2^{+12.7}_{-9.5}\pm1.4$ & $5.9^{+1.8}_{-1.3}\pm0.2$\\	
  2.6444	  & 33.7 	      & 16.7  & $33.1\pm7.7$  & $5.8\pm1.4\pm0.4$ & $2.8\pm0.3\pm0.1$\\
  2.6464	& 34.0		      & 16.8  &$38.0\pm8.4$  & $6.6\pm1.5\pm0.5$ & $2.9\pm0.3\pm0.1$\\
  2.9000   	& 105.	     & 14.2 & $18.0\pm7.1$ & $1.2\pm0.5\pm0.1$ & $1.2\pm0.2\pm0.1$\\
  *2.9884	& 65.2		   &14.9 & $9.4^{+5.4}_{-4.6}$ & $1.0^{+0.6}_{-0.5}\pm0.1$ & $1.1\pm0.3\pm0.1$\\
  \hline
  \end{tabular}
  \end{center}
  \end{table*}

\begin{multicols}{2}
Systematic uncertainties associated with the cross section measurements include event selection, 
cross section line-shape, angular distribution, fitting method, energy scale, and luminosity.  
In the nominal results, the differences  of data and MC efficiencies are corrected with control samples.
We vary the data/MC correction factors within their $\pm 1\sigma$ uncertainty 
and the resulting  differences in the cross sections are taken as the uncertainty from the event selection.
The uncertainty associated with the cross section line-shape 
is 1.0\%, which includes both the theoretical uncertainty and the parameter uncertainty in the line-shape fit.
The uncertainty from the angular distribution is evaluated by varying $|G_{E}/G_{M}|$ ratios within $\pm 1\sigma$ at the
three energy points with the highest statistics. For the energy points with 
unknown $|G_{E}/G_{M}|$ values, two extreme cases
$G_{E}=0$ and $G_{M}=0$ are considered and the difference in the efficiencies divided by
a factor of $\sqrt{12}$ is taken as the uncertainty~\cite{systematic}.
Alternative fits are performed to study the uncertainty from the fit procedure. These include varying the fitting range, varying the signal
shape by fixing the resolution of the convolved Gaussian to be $\pm1\sigma$ different from its nominal value,
and changing the background PDF from a second order to a third order polynomial.
The effects of the c.m.~energy and energy resolution uncertainties are studied for energy points near threshold.
The difference of the cross sections in  $e^{+}e^{-}\to\Sigma^{+}\bar{\Sigma}^{-}$ is very small and the corresponding uncertainty on
the cross sections can be neglected.
The uncertainty on the effective FFs  are 4.9\% and 2.8\%  at $\sqrt{s}=2.3864$ and 2.396 GeV due to the change
of Coulomb correction factors.
For the $e^{+}e^{-}\to\Sigma^{-}\bar{\Sigma}^{+}$ process, the variation of c.m~energy and energy resolution introduce 
uncertainties of 12.0\%
and 14.2\% in the cross section and effective FF, respectively, at $\sqrt{s}=2.396$~GeV.
The integrated luminosity is determined with large angle Bhabha events with an uncertainty of 1.0\%~\cite{dataset}.
All sources of systematic uncertainties are treated as uncorrelated and summed in quadrature;
they are in the range between  3.5\% and 13.0\% of the cross sections, depending on the c.m.~energy.

\section{Line shape analysis}
\begin{figure}[H] 
\begin{center}
\begin{overpic}[width=6.5cm,height=4.5cm,angle=0]{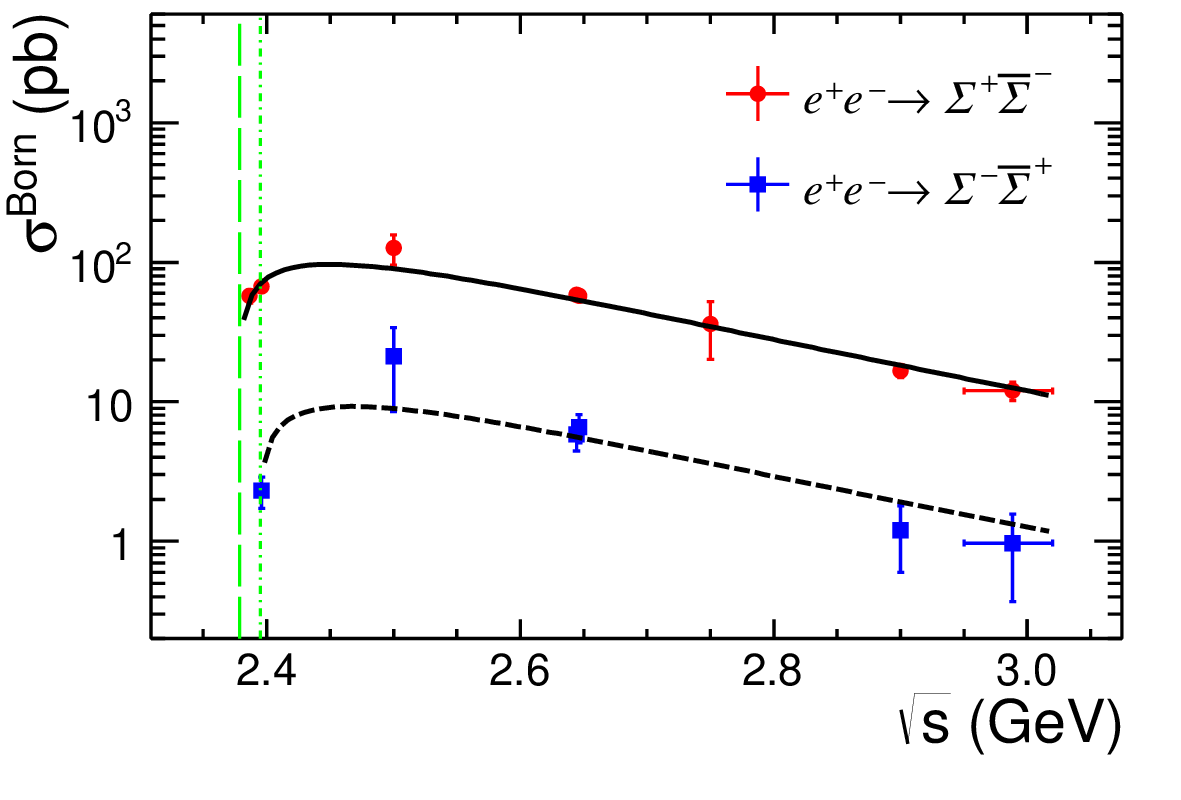}
\end{overpic}
\caption{ (color online) The cross section lineshapes for $e^{+}e^{-}\to\Sigma^{+}\bar{\Sigma}^{-}$ (circles) and $e^{+}e^{-}\to\Sigma^{-}\bar{\Sigma}^{+}$ (squares).
 The solid line is the pQCD fit for $e^{+}e^{-}\to\Sigma^{+}\bar{\Sigma}^{-}$  and the dashed line for $e^{+}e^{-}\to\Sigma^{-}\bar{\Sigma}^{+}$.
The vertical dashed and dotted lines denote production thresholds for $e^{+}e^{-}\to\Sigma^{+}\bar{\Sigma}^{-}$ and $e^{+}e^{-}\to\Sigma^{-}\bar{\Sigma}^{+}$.}
\label{comgeff}
\end{center}
\end{figure}

The measured cross section line-shapes of $e^{+}e^{-}\to\Sigma^{\pm}\bar{\Sigma}^{\mp}$
from $\sqrt{s}=2.3864$ to 3.0200~GeV
are shown in Fig.~\ref{comgeff}. 
The near threshold cross sections for $e^{+}e^{-}\to\Sigma^{+}\bar{\Sigma}^{-}$
and  $e^{+}e^{-}\to\Sigma^{-}\bar{\Sigma}^{+}$ are measured to be $58.2\pm5.9^{+2.8}_{-2.6}$
and $2.3\pm0.5\pm0.3$~pb, respectively, 
both are inconsistent with the value of 520~pb expected for point-like charged baryons.
Instead, a new feature is observed in which 
the cross sections for $e^{+}e^{-}\to\Sigma^{-}\bar{\Sigma}^{+}$ are consistently
smaller than those for $e^{+}e^{-}\to\Sigma^{+}\bar{\Sigma}^{-}$.
A perturbative QCD-motivated energy power function~\cite{rinaldo,QCD}, given by
\begin{equation}
\label{pqcd}
\sigma^{\text{B}}(s) = \frac{\beta C}{s}\left(1+\frac{2m_{B}^{2}}{s}\right)\frac{c_{0}}{(s-c_{1})^{4} [\pi^{2}+\text{ln}^{2}(s/\Lambda_{\text{QCD}}^{2})]^{2}}
\end{equation}
is used to fit the line-shapes, where $c_{0}$ is the normalization, $c_{1}$ is the mean effect of a set of intermediate states that mediates the coupling between the virtual photon~\cite{VMD}
 and is regarded as common for the two processes, and $\Lambda_{\text{QCD}}$ is the QCD scale, fixed to 0.3~GeV. 
The fit results are shown in Fig.~\ref{comgeff} with a fit quality of $\chi^{2}/{\rm ndof}=9.7/12$,
where $\rm ndof$ is number of degrees of freedom.
The cross section ratio between $e^{+}e^{-}\to\Sigma^{+}\bar{\Sigma}^{-}$ and $e^{+}e^{-}\to\Sigma^{-}\bar{\Sigma}^{+}$ is obtained from $c_{0}$
to be $9.7\pm1.3$, and $c_{1}$ is $2.0\pm0.2$~GeV$^{2}$.
Since the effective FF is proportional to the square root of the Born cross section,
the ratio of the effective $\Sigma^{+}$ and $\Sigma^{-}$ FFs is consistent with 3,
which is the ratio of the incoherent sum of the squared charges of the $\Sigma^{+}$ and $\Sigma^{-}$ valence quarks, $\sum_{\text q \in \text B}Q_{q}^{2}$.

The results are in disagreement with the prediction from octet baryon wave functions~\cite{ratio_theory2}, where
the typical $SU(3)$-symmetry breaking effects for hyperon FFs are about $10\sim30\%$. 
In the di-quark model, the $\Sigma^{+}$ FFs should be comparable to that of  $\Lambda$~\cite{diquark}.
The $\Sigma^{\pm}$ FFs are also predicted in Ref.~\cite{UandA} from Unitary and Analytic model.
We notice that a recent prediction for the non-resonant cross section of $e^{+}e^{-}\to \Sigma^{\pm}\bar{\Sigma}^{\mp}$ at the $J/\psi$ mass~\cite{rinaldo2}, based on an effective Lagrangian density,
 is consistent with our result when extrapolated to $\sqrt{s}=3.097$~GeV using Eq.~(\ref{pqcd}).

\section{Extraction of $|G_{E}/G_{M}|$ ratio}
\begin{figure}[H] 
\begin{center}
\begin{overpic}[width=8.cm,height=3.5cm,angle=0]{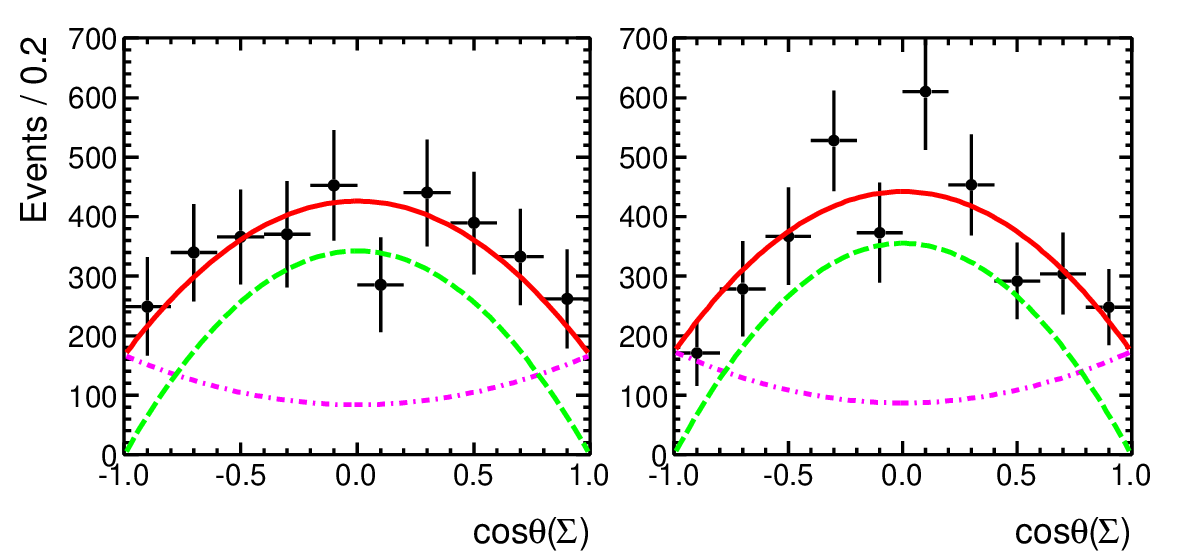}
\put(15, 32){\small{a)}}
\put(62,32){\small{b)}}
\end{overpic}
\caption{(color online) Simultaneous fit of efficiency corrected angular distribution at $\sqrt{s}=2.396$~GeV
for a) category A b) category B for $e^{+}e^{-}\to\Sigma^{+}\bar{\Sigma}^{-}$ events.
Dots with error bars are data, solid curves are the fit results, the contributions from $G_{E}$
and $G_{M}$ are indicated by dashed and dotted curves.}
\vspace{-0.5cm}
\label{figangular2396}
\end{center}
\end{figure}

The value of $|G_{E}/G_{M}|$ can be obtained by fitting the differential
angular distribution according to Eq.~(\ref{equ:eq1}). The statistics
at $\sqrt{s}=2.3960$, 2.6444, 2.6464 and 2.9000~GeV for
 $e^{+}e^{-}\to\Sigma^{+}\bar{\Sigma}^{-}$ allow us to perform a study of the
polar angle of $\Sigma^{+}$ in the c.m.~frame.
The  angular distributions for categories~A and~B at $\sqrt{s}=2.3960$~GeV
are shown in Fig.~\ref{figangular2396}.
These angular distributions have been corrected for the detection efficiency and ISR, 
which are obtained from signal MC simulation. Additional bin-by-bin corrections
 due to the data/MC detection differences,
for categories~A and~B, respectively, have also been applied. 
Simultaneous fits to the two data sets
to the expression in Eq.~(\ref{equ:eq1}) sharing a common value for $|G_{E}/G_{M}|$ are performed. 
The result of $|G_{E}/G_{M}|=1.83\pm 0.26$ is significantly higher than 1.
Using the normalized number of events, $|G_{M}|$ is determined to be
$(9.14\pm 1.42)\times10^{-2}$ and
 $(9.30\pm 1.53)\times10^{-2}$ for category A and B, respectively.
Similar angular distribution fits are performed for the combined $\sqrt{s}=2.6444$
and 2.6464~GeV data sets, denoted as 2.6454~GeV, and  $\sqrt{s}=2.90$~GeV and the results are listed in Table~\ref{sumsigmap}.
The systematic uncertainties on $|G_{E}/G_{M}|$ considered here are the difference between data and MC efficiency, the bin size, and the
fit range. 
For the $\Sigma^{-}$, on the other hand, the statistics only allow for the determination of $|G_{\rm eff}|$; they are not sufficient to extract $|G_{E}/G_{M}|$.

\section{Summary}

In summary, the data collected by BESIII at c.m.~energies between $2.3864$ and 3.0200~GeV,
are exploited to perform measurements of $e^{+}e^{-}\to\Sigma^{\pm}\bar{\Sigma}^{\mp}$.
This is the first time that cross sections of $e^{+}e^{-}\to\Sigma^{\pm}\bar{\Sigma}^{\mp}$ 
in the off-resonance region are presented.
The precision has been significantly improved by reconstructing all dominant decay modes of the $\Sigma$. 
Cross sections near threshold are observed for 
$e^{+}e^{-}\to\Sigma^{+}\bar{\Sigma}^{-}$ and $e^{+}e^{-}\to\Sigma^{-}\bar{\Sigma}^{+}$
to be $58.2\pm5.9^{+2.8}_{-2.6}$ and $2.3\pm0.5\pm0.3$~pb, respectively. 
The values disagree with the point-like expectations near threshold, $848(m_{p}/m_{B})^{2}$~pb, as has been seen for the proton~\cite{time_babar,time_cmd}. 
The cross section line-shapes for $e^{+}e^{-}\to\Sigma^{+}\bar{\Sigma}^{-}$ and
$e^{+}e^{-}\to\Sigma^{-}\bar{\Sigma}^{+}$ are well-described by pQCD-motivated functions.
The ratio of the $\sigma^{\text{Born}}(e^{+}e^{-}\to\Sigma^{+}\bar{\Sigma}^{-})$ to
$\sigma^{\text{Born}}(e^{+}e^{-}\to\Sigma^{-}\bar{\Sigma}^{+})$ is determined to be $9.7\pm1.3$,
which is inconsistent with predictions from various models~\cite{ratio_theory2, UandA, diquark}. The EMFF ratio $|G_{E}/G_{M}|$ of the $\Sigma^{+}$ is determined from its production angle dependence
at three high-statistics energy points.
The $|G_{E}/G_{M}|$ of the $\Sigma^{+}$ shows similar features to those of the proton~\cite{time_babar,time_bes3}, $\Lambda$~\cite{hyperon_bes32},
 and $\Lambda_{c}$~\cite{lambdac_bes3},
that is larger than 1 within uncertainties near threshold
 and consistent with 1 at higher c.m.~energies. \\

\section*{Acknowledgements}
The BESIII collaboration thanks the staff of BEPCII and the IHEP computing center and the supercomputing center of USTC for their strong support. This work is supported in part by National Key Research and Development Program of China under Contracts Nos. 2020YFA0406300, 2020YFA0406400; National Natural Science Foundation of China (NSFC) under Contracts Nos.~11625523, 11635010, 11605196, 11605198, 11705192, 11735014, 11822506, 11835012, 11935015, 11935016, 11935018, 11961141012, 12022510, 12035013, 11950410506, 12061131003; the Chinese Academy of Sciences (CAS) Large-Scale Scientific Facility Program; Joint Large-Scale Scientific Facility Funds of the NSFC and CAS under Contracts Nos. U1732263, U1832103, U1832207, U2032111; CAS Key Research Program of Frontier Sciences under Contract No. QYZDJ-SSW-SLH040; 100 Talents Program of CAS; INPAC and Shanghai Key Laboratory for Particle Physics and Cosmology; ERC under Contract No. 758462; European Union Horizon 2020 research and innovation programme under Contract No. Marie Sklodowska-Curie grant agreement No 894790; German Research Foundation DFG under Contracts Nos. 443159800, Collaborative Research Center CRC 1044, FOR 2359, FOR 2359, GRK 214; Istituto Nazionale di Fisica Nucleare, Italy; Ministry of Development of Turkey under Contract No. DPT2006K-120470; National Science and Technology fund; Olle Engkvist Foundation under Contract No. 200-0605; STFC (United Kingdom); The Knut and Alice Wallenberg Foundation (Sweden) under Contract No. 2016.0157; The Royal Society, UK under Contracts Nos. DH140054, DH160214; The Swedish Research Council; U. S. Department of Energy under Contracts Nos. DE-FG02-05ER41374, DE-SC-0012069

\end{multicols}


\begin{thebibliography}{99}
\bibitem{stern} R.~Frisch, O.~Stern, Z. Physik {\bf 85}, 4 (1933).

\bibitem{hofstadter} R.~Hofstadter, Rev. Mod.~Phys. {\bf 28}, 214 (1956).

\bibitem{protonradius} J.~C.~Bernauer {\it et al.} Phys.\ Rev.\ Lett. {\bf 105}, 242001 (2010). 
R.~Pohl {\it et al.}, Nature {\bf 466}, 213 (2010).
N.~Bezginov {\it et al.} Science {\bf 365}, no. 6457, 1007 (2019).


\bibitem{hyperon1} G.~Eichmann, H.~Sanchis-Alepuz, R.~Williams, R.~Alkofer and C.~S.~Fischer, Prog.\ Part.\ Nucl.\ Phys.\  {\bf 91}, 1 (2016).

\bibitem{hyperon2} G.~Ramalho, K.~Tsushima and A.~W.~Thomas,  J.\ Phys.\ G {\bf 40}, 015102 (2013);
F.~Gross, G.~Ramalho and K.~Tsushima,  Phys.\ Lett.\ B {\bf 690}, 183 (2010).

\bibitem{chiral} L.~S.~Geng, J.~Martin Camalich, L.~Alvarez-Ruso and M.~J.~Vicente Vacas, Phys.\ Rev.\ Lett.\  {\bf 101}, 222002 (2008).
\bibitem{pqcd}  S.~J.~Brodsky and G.~R.~Farrar, Phys.\ Rev.\ D {\bf 11}, 1309 (1975).
\bibitem{lattice1}  J.~R.~Green, J.~W.~Negele, A.~V.~Pochinsky, S.~N.~Syritsyn, M.~Engelhardt and S.~Krieg, Phys.\ Rev.\ D {\bf 90}, 074507 (2014). 


\bibitem{space_ros} I.~A.~Qattan {\it et al.},  Phys.\ Rev.\ Lett.\  {\bf 94}, 142301 (2005).
\bibitem{space_jlab} A.~J.~R.~Puckett {\it et al.}, Phys.\ Rev.\ C {\bf 85}, 045203 (2012).

\bibitem{time_cleo}  T.~K.~Pedlar {\it et al.} [CLEO Collaboration], Phys.\ Rev.\ Lett.\  {\bf 95}, 261803 (2005).
\bibitem{time_babar} J.~P.~Lees {\it et al.} [BaBar Collaboration], Phys.\ Rev.\ D {\bf 87}, 092005 (2013).
\bibitem{time_cmd} R.~R.~Akhmetshin {\it et al.} [CMD-3 Collaboration], Phys.\ Lett.\ B {\bf 759}, 634 (2016).
\bibitem{time_bes3}  M.~Ablikim {\it et al.} [BESIII Collaboration], Phys.\ Rev.\ Lett.\  {\bf 124}, 042001 (2020).

\bibitem{hyperon_babar} B.~Aubert {\it et al.} [BaBar Collaboration], Phys.\ Rev.\ D {\bf 76}, 092006 (2007).
\bibitem{hyperon_cleo} S.~Dobbs, K.~K.~Seth, A.~Tomaradze, T.~Xiao and G.~Bonvicini, Phys.\ Rev.\ D {\bf 96}, 092004 (2017).
\bibitem{hyperon_bes3}  M.~Ablikim {\it et al.} [BESIII Collaboration], Phys.\ Rev.\ D {\bf 97}, 032013 (2018).
\bibitem{hyperon_bes32} M.~Ablikim {\it et al.} [BESIII Collaboration], Phys.\ Rev.\ Lett.\  {\bf 123}, 122003 (2019)

\bibitem{nnbar_fenice} A.~Antonelli {\it et al.},  Nucl.\ Phys.\ B {\bf 517}, 3 (1998).
\bibitem{nnbar_bes3} S.~Ahmed for the BESIII Collaboration, ``New features of the neutron electromagnetic structure in the electron-positron annihilation", talk at Lake Louise Winter Institute 2020, Feb.~9-15, Alberta, Canada.



\bibitem{ratio_theory1}   J.~R.~Ellis and M.~Karliner, New J.\ Phys.\  {\bf 4}, 18 (2002) 
\bibitem{ratio_theory2}  V.~L.~Chernyak and A.~R.~Zhitnitsky, Phys.\ Rept.\  {\bf 112}, 173 (1984);
V.~L.~Chernyak, A.~A.~Ogloblin and I.~R.~Zhitnitsky,  Z.\ Phys.\ C {\bf 42}, 569 (1989).





\bibitem{diquark}   
M. Anselmino {\it et al.}, Rev. Mod. Phys. {\bf 65}, 1199 (1993);
R.~L.~Jaffe and F.~Wilczek, Phys.\ Rev.\ Lett.\  {\bf 91}, 232003 (2003);
 R.~L.~Jaffe, Phys.\ Rept.\  {\bf 409}, 1 (2005)

\bibitem{UandA}  G.~Ramalho, M.~T.~Pena, and K.~Tsushima, Phys.\ Rev.\ D {\bf 101}, 014014 (2020).

\bibitem{ee} N.~Cabibbo and R.~Gatto. Phys. Rev. {\bf 124}, 1577 (1961).
\bibitem{schwinger} J.~Schwinger, $Particle,~Sources,~and~Field$~(Perseus Books Publishing, Massachusetts, 1998), Vol.3.
\bibitem{coulomb}  A.~B.~Arbuzov and T.~V.~Kopylova, JHEP {\bf 1204}, 009 (2012).



\bibitem{pdg} P.~A.~Zyla {\it et al.} [Particle Data Group], Prog.\ Theor.\ Exp. Phys. {\bf 2020}, 083C01 (2020).
\bibitem{lambdac_bes3}  M.~Ablikim {\it et al.} [BESIII Collaboration], Phys.\ Rev.\ Lett.\  {\bf 120}, 132001 (2018).

\bibitem{int_fsi}   L.~Y.~Dai, J.~Haidenbauer and U.~G.~Meissner, Phys.\ Rev.\ D {\bf 96}, 116001 (2017).
\bibitem{int_res}  B.~El-Bennich, M.~Lacombe, B.~Loiseau and S.~Wycech, Phys.\ Rev.\ C {\bf 79}, 054001 (2009).
\bibitem{int_coulomb}  R.~Baldini~Ferroli, S.~Pacetti, A.~Zallo and A.~Zichichi, Eur.\ Phys.\ J.\ A {\bf 39}, 315 (2009);
R.~Baldini~Ferroli, S.~Pacetti and A.~Zallo, Eur.\ Phys.\ J.\ A {\bf 48}, 33 (2012).




\bibitem{dataset} M.~Ablikim {\it et al.} [BESIII Collaboration], Chin.\ Phy.\ C {\bf 41}, 063001 (2017).
\bibitem{Ablikim:2009aa} M.~Ablikim {\it et al.} [BESIII Collaboration], Nucl.\ Instrum.\ Meth.\ A {\bf 614}, 345 (2010).
\bibitem{geant4} S.~Agostinelli {\it et al.} [GEANT4 Collaboration], Nucl.\ Instrum.\ Meth.\ A {\bf 506}, 250 (2003).

\bibitem{Anjrej}
G.~F\"aldt, Eur. Phys.~J.~A {\bf 51}, 74 (2015);
G.~F\"aldt, Eur. Phys. J. A {\bf 52}, 141 (2016).


\bibitem{conexc} R. G. Ping, Chin.\ Phys.\ C {\bf 38}, 083001 (2014).
\bibitem{babayaga} G.~Balossini {\it et al.}, Nucl.\ Phys.\ B {\bf 758}, 227 (2006);
G.~Balossini {\it et al.}, Phys.\ Lett.\ B {\bf 663}, 209 (2008).
\bibitem{lundarlw} B.~Andersson and H.~Hu, hep-ph/9910285.
\bibitem{bestwogam}  S.~Nova, A.~Olchevski and T.~Todorov, DELPHI-90-35 PROG 152 (1990).
S. Nova, A.~Olchevski and T.~Todorov [DELPHI Collaboration], DELPHI 90-35 PROG {\bf 152} 1990.


\bibitem{pinglamlam} M.~Ablikim {\it et al.} [BESIII Collaboration], Phys. Rev. D \textbf{101}, 092002 (2020).

\bibitem{VP} S.~Actis {\it et al.} [Working Group on Radiative Corrections and Monte Carlo Generators for Low Energies],
  Eur.\ Phys.\ J.\ C {\bf 66}, 585 (2010)

\bibitem{argus} H.~Albrecht {\it et al.} [ARGUS Collaboration], Phys.\ Lett.\ B {\bf 241}, 278 (1990).
\bibitem{effectiveff} M.~Ablikim {\it et al.} [BESIII Collaboration], Phys.\ Rev.\ D {\bf 91}, 112004 (2015).





\bibitem{systematic} R.~Wanke, "Systematic Uncertainties", Terascale Statistics Tools School, DESY, Mar 21, 2015-p.1/56.

\bibitem{rinaldo} S.~Pacetti, R.~Baldini~Ferroli and E.~Tomasi Gustafsson, Phys. Rept. {\bf 550-551}, 1 (2015).

\bibitem{QCD} G.~P.~Lepage and S.~J.~Brodsky,
Phys. Rev. D \textbf{22}, 2157 (1980).

\bibitem{VMD} H.~W.~Hammer, U.~G.~Meissner and D.~Drechsel,
  Phys.\ Lett.\ B {\bf 385}, 343 (1996).

\bibitem{rinaldo2} R.~Baldini Ferroli, A.~Mangoni, S.~Pacetti and K.~Zhu,
  Phys.\ Lett.\ B {\bf 799}, 135041 (2019).


\end{thebibliography}
\end{document}